\documentclass[12pt,a4paper]{article}
\usepackage[utf8]{inputenc}
\usepackage{amsmath}
\usepackage{amsfonts}
\usepackage{amssymb}
\usepackage{graphicx}
\numberwithin{equation}{section}
\usepackage{appendix}
\usepackage{srcltx}
\usepackage{authblk}
\usepackage{cancel}
\usepackage[left=2.00cm, right=2.00cm, top=2.00cm, bottom=2.00cm]{geometry}

\newcount\Comments  %
\makeatletter

\begin{document}
\title{Coalescence, Deformation and B\"acklund  Symmetries 
of Painlev\'e IV and II Equations}
\author[1]{V.C.C. Alves}
\author[2]{ H. Aratyn}
\author[1]{J.F. Gomes} 
\author[1]{A.H. Zimerman}
 \affil[1]{
 Instituto de F\'{\i}sica Te\'{o}rica-UNESP,
 Rua Dr Bento Teobaldo Ferraz 271, Bloco II,
 01140-070 S\~{a}o Paulo, Brazil}
\affil[2]{
Department of Physics, 
 University of Illinois at Chicago,
 845 W. Taylor St.
 Chicago, Illinois 60607-7059}

\maketitle

\abstract{
{We extend  Painlev\'e IV model by adding quadratic terms to its Hamiltonian
obtaining two classes of %
models (coalescence and deformation) that interpolate between Painlev\'e IV and II equations for special 
limits of the underlying parameters.}
We derive the underlying B\"acklund transformations, symmetry structure and
requirements to satisfy Painlev\'e
property.

}

\section{Introduction}

The Painlev\'e equations are second-order 
differential equations whose solutions have no movable singular 
points except poles. 
This feature (pure poles are the only movable singularities)
of some second order differential equations is known as Painlev\'e property.  
The Painlev\'e equations naturally 
emerge as special scaling limits  of integrable models 
\cite{flaschka,p4aip,higherpainleve,anpz,p6}
and a fundamental conjecture  \cite{ars} establishes connection between  
Painlev\'e property and solvability by inverse scattering.
Another basic aspect of Painlev\'e equations and their Hamiltonian
structures is invariance under extended affine Weyl symmetry groups
\cite{higher,noumi}. For example the fourth  Painlev\'e  equation, to which 
we will refer as P$_{\rm IV}$, exhibits symmetry under B\"acklund 
transformations that form the affine Weyl group of type $A^{(1)}_2$ 
and the second Painlev\'e  equation, to which we will refer as P$_{\rm II}$, 
is invariant under B\"acklund transformations from 
the affine Weyl group $A^{(1)}_1$. {B\"acklund transformations have 
also been  extensively studied in connection with the Schlesinger 
transformations, see for instance  references \cite{Fokas,Mugan,Ustinov}
for the case of Painleve  II and IV equations. }

Hybrid Painlev\'e equations have been a focus of several  papers, e.g. 
\cite{kudryashov,rogers}. 
More recently, in reference \cite{p3-p5} we introduced the hybrid P$_{\rm III-V}$ model 
that was obtained as reduction of a class
of integrable models known as multi-boson systems
\cite{higherpainleve,anpz} 
that generalize the AKNS hierarchy \cite{p4aip}. 
The P$_{\rm III-V}$ model reduces 
to P$_{\rm III}$,  P$_{\rm V}$ and I$_{\rm 12}$, I$_{\rm 38}$ and I$_{\rm 49}$ 
equations from Ince's list \cite{ince,inceshams}  for 
special limits of its parameters while  for remaining finite
values of its parameters preserves enough symmetry under remaining B\"acklund transformations of the 
extended affine Weyl symmetry group to satisfy Painlev\'e property
\cite{p3-p5}. 
 
We will conduct here a similar
investigation for the hybrid %
of P$_{\rm II}$ and P$_{\rm IV}$ models and point out how the presence 
of remaining B\"acklund transformations symmetries influences
the outcome of the  Painlev\'e test.
Starting from  the  symmetric Painlev\'e IV equations, in section 
\ref{section:P4model},
we enlarge its parameter space to allow for extension of 
symmetry structure by additional automorphisms $\pi_i,
\rho_i, i=0,1,2$. %
We derive algebraic relations between these automorphisms
and $A^{(1)}_2$ B\"acklund transformations. 

We present two different limiting procedures leading to Painlev\'e II equation.

One way, described in section \ref{section:coalescence}, 
is to formulate coalescence/degeneracy 
in a framework of symmetric Painlev\'e IV equations augmented by
a non-zero integration constant. This generalization of P$_{\rm IV}$
equation %
remains invariant under the additional automorphism $\rho_2$.
The underlying Weyl group
symmetry reduces  from  $A^{(1)}_2$  down 
to $A^{(1)}_1$ in the appropriate limit and we are able to obtain 
close expressions for the B\"acklund transformations  of P$_{\rm II}$
from their P$_{\rm IV}$ counterparts.
In the  P$_{\rm II}$ limit  the automorphism $\rho_2$ toggles between
two copies of  P$_{\rm II}$ equations each with its own $A^{(1)}_1$
symmetry.

In another scheme, presented in section \ref{section:deformation}, 
the $A^{(1)}_2$ symmetry group of symmetric
Painlev\'e IV equation is explicitly broken by addition of a 
deformation parameter before the limit %
resulting in Painlev\'e II equation is taken. 
The deformed model is
formulated in such a way that it is invariant under additional
automorphisms $\pi_2, \rho_2$.
We point out a
connection between existence of residual symmetry of the deformed model 
(invariance under one of the original three B\"acklund transformations
of $A^{(1)}_2$)
and passing of the Kovalevskaya-Painlev\'e test by this model.
Such deformed model %
provides another example of hybrid Painlev\'e equations 
with properties that they pass Painlev\'e test, retain invariance under residual  
B\"acklund transformations and reduce down to 
underlying Painlev\'e or Ince equations for special values of their parameters.

{
In section \ref{section:hamdeform} we will introduce and study 
a generalization of P$_{\rm IV}$  Hamiltonian structure 
of the form :
\begin{equation}
H= H_0
 +\frac{1}{\epsilon} (f_0+f_1) \left( k_1 \sigma z -
\frac{k_2}{2} (f_0+f_1)  \right)\,,
\label{hamilton}
\end{equation}
where $\epsilon, \sigma, k_1, k_2$ are complex parameters 
and 
\begin{equation}
H_0= -f_0f_1f_2 +
 \frac{-\alpha_1+\alpha_2}{3}f_0+\frac{-\alpha_1-2\alpha_2}{3}f_1 +
 \frac{2\alpha_1+\alpha_2}{3}f_2\, {,}
\label{okamotoP4}
\end{equation}
is the well-known Okamoto's  P$_{\rm IV}$  Hamiltonian \cite{okamoto4}.
The two basic conditions that guide our construction of such generalization are :
(1) that the original cubic Hamiltonian
is augmented only by  terms of dimensions lower than 
three and (2) the Hamilton equations remain finite and
do not violate the  Painlev\'e property.
These conditions restrict the allowed generalization
of P$_{\rm IV}$  Hamiltonian structure to be of the form given 
in equation \eqref{hamilton}.
}
As we will see below the combination $f_0+f_1$ appearing in the above expression ensures
invariance under a pair of B\"acklund transformations $s_2,\rho_2$, if
we used $f_0+f_2$ or $ f_1+f_2$ we would encounter invariance under 
$s_1,\rho_1$ or $s_0,\rho_0$ with all these transformations being
defined in the forthcomming sections.

We show that this natural generalization \eqref{hamilton} 
represents either coalescence/degeneracy
or $A^{(1)}_2$ deformation of P$_{\rm IV}$ and we present arguments that those 
two approaches are the only  ones leading  from P$_{\rm IV}$ 
model to P$_{\rm II}$ model under the above conditions.

{We summarize the novel features of our 
formalism and reiterate rationale for expanding the parameter space of
Painlev\'e IV model by additional parameters in Section
\ref{section:concluding}.}

\section{The structure of  P$_{\rm IV}$ model, definition and symmetries}
\label{section:P4model}
This section is devoted to a summary of relevant results on P$_{\rm IV}$ 
equations, B\"acklund transformations and coalescence between
P$_{\rm IV}$  and P$_{\rm II}$ available in 
the literature (e.g. \cite{gromak,noumi}). 

We also generalize the conventional symmetric Painlev\'e IV model 
by adding the new parameter $\sigma$ in a way that makes the generalized
model invariant under additional automorphisms $\pi_i, \rho_i, \, i=0,1,2$
satisfying the braid relations.

\subsection{P$_{\rm IV}$ symmetric equations}
The starting point of subsection is the Okamoto Hamiltonian 
\eqref{okamotoP4} for P$_{\rm IV}$ equation. In the literature
the parameters $\alpha_i, i=0,1,2$ 
satisfy the condition $\alpha_1+\alpha_2+\alpha_0=1$.
Here we find that our discussion of symmetries and coalescence
limits will profit from working instead 
with conditions :
\begin{equation}
\alpha_1+\alpha_2+\alpha_0=\sigma , \quad f_2=\sigma z-f_1-f_0\, {.}
\label{alpha0f0}
\end{equation}
Here we introduced $\sigma$ as an additional parameter
for the  P$_{\rm IV}$ model that enables us to extend 
symmetry group of the model. 
{The advantages of introducing the $\sigma$ parameter will be summarized in
the concluding Section \ref{section:concluding}.}

The corresponding Hamilton's equations 
can be cast in a form of the so-called
symmetric P$_{\rm IV}$
system described by e.g. \cite{noumi}:
\begin{equation}
\begin{split}
f_0'&=f_0 \left(f_1-f_2\right)+\alpha_0\, ,\\
f_1'&=f_1 \left(f_2-f_0\right)+\alpha_1\, ,\\
f_2'&=f_2\left(f_0-f_1\right)+\alpha_2\, ,
\label{symp4}
\end{split}
\end{equation}
where $f_i = f_i(z)$ and $'=d/dz$.

Eliminating $f_2=\sigma z-f_0-f_1$ from \eqref{symp4} we obtain:
\begin{equation}
\begin{split}\label{noumisum}
f_0'(z)=&f_0 \left(-\sigma z+f_0+2 f_1\right)+\alpha _0\,,\\
f_1'(z)=&f_1\left(\sigma z-2 f_0-f_1\right)+\alpha _1\,, 
\end{split}
\end{equation}
while the third equation in \eqref{symp4} can be obtained
by summing the above two equations.

By further eliminating $f_1$ or $f_0$ from \eqref{noumisum}  we get
for the remaining component:
\begin{equation}\label{p4}
f_i''(z)=\frac{f_i'{}^2}{2 f_i}-\frac{\alpha _i^2}{2 f_i}
+\left(\frac{1}{2}\sigma^2 z^2+  {(-1)^i(2 \alpha _0+2 \alpha
_1-\alpha_i- \sigma)}\right) f_i-2 \sigma  z f_i{}^2
+\frac{3}{2} f_i{}^3
,\quad \; i=0,1.
\end{equation}
{Both equations are} equivalent to the standard P$_{\rm IV}$ equation
\cite{gromak,p4aip}:
\begin{equation}
w_{xx}=\frac{{w_{x}}^2}{2w}+\frac{3w^3}{2}+4x w^2+
2\left(x^2-A \right)w+\frac{B }{w}
\end{equation}
by setting $\sigma\to1$ followed by transformations
\begin{equation}\label{p4a}
f_0(z)=\frac{w(x)}{\sqrt{-2}},\qquad z=x\sqrt{-2}\qquad 
\alpha _1= \frac{1}{2} \left(1+A-\alpha _0\right),\qquad
\alpha _0= \sqrt{\frac{-B}{2}}
\end{equation}
and a similar transformation for $f_1$ with the appropriate changes.

Equations \eqref{p4} %
will be referred to as P$_{\rm
IV}$ equations throughout this document while equations \eqref{symp4} will be
referred to as symmetric P$_{\rm IV}$ equations.

\subsection{B\"acklund and auto-B\"acklund Transformations}\label{BTsecond}

Equations \eqref{symp4} are manifestly invariant under B\"acklund transformations
$s_i$ ($i=0,1,2$) and automorphism $\pi$ defined  as follows {(see
e.g. \cite{noumi})}:
\begin{equation}
\begin{array}{c|ccc|ccc|ccc}
{} & {\alpha_0} & { \alpha_1} &
{ \alpha_2} & { f_0} & { f_1} &
{ f_2}\\
\hline
{ s_0}&{ -\alpha_0} &{
	\alpha_1+\alpha_0} &{ \alpha_2+\alpha_0} &{
	f_0} &{ f_1+\frac{\alpha_0}{f_0}} &{
	f_2-\frac{\alpha_0}{f_0}}\\[2mm]
\hline
{ s_1}&{ \alpha_0+\alpha_1} &{
	-\alpha_1} &{ \alpha_2+\alpha_1} &{
	f_0-\frac{\alpha_1}{f_1}} &{ f_1} &{
	f_2+\frac{\alpha_1}{f_1}} \\[2mm]
\hline
{ s_2}&{ \alpha_0+\alpha_2} &{
	\alpha_1+\alpha_2} &{-\alpha_2} &{
	f_0+\frac{\alpha_2}{f_2}} &{ f_1-\frac{\alpha_2}{f_2}} &{
	f_2} \\[2mm]
\hline
{ \pi}&{ \alpha_1} &{
	\alpha_2} &{ \alpha_0} &{
	f_1} &{ f_2} &{ f_0} 
\end{array}
\label{p4symm}
\end{equation}

These transformations satisfy 
\begin{equation}
s_i^2=1,\quad (s_is_{i+1})^3=1,\quad \pi^3=1,\quad \pi
s_i=s_{i+1}\pi,\quad i =0,1,2 \, \cancel{{,}},
\end{equation}
and thus $\langle s_0,s_1,s_2,\pi\rangle$ form the extended affine Weyl group $\mathcal{A}_2^{(1)}$ \cite{noumi}.

Due to the presence of parameter $\sigma$ introduced in equation 
\eqref{alpha0f0}
in the setting of symmetric P$_{\rm IV}$ equation  \eqref{symp4} 
we have additional automorphisms $\pi_i$ and $ \rho_i, i=0,1,2$  :
  \begin{equation}
\begin{array}{c|ccc|ccc|c|}
{} & {\alpha_0} & { \alpha_1} &
{ \alpha_2} & { f_0} & { f_1} &
{ f_2}& \sigma \\
\hline
{ \pi_0}&{ -\alpha_0} &{
 -\alpha_2} &{ -\alpha_1} &{
 -f_0} &{ -f_2} &{ -f_1}&-\sigma \\
\hline
{ \pi_1}&{ -\alpha_2} &{
 -\alpha_1} &{ -\alpha_0} &{
 -f_2} &{ -f_1} &{ -f_0}&-\sigma \\
\hline
{ \pi_2}&{ -\alpha_1} &{
 -\alpha_0} &{ -\alpha_2} &{
 -f_1} &{ -f_0} &{ -f_2}&-\sigma  
\end{array}
\label{piss}
\end{equation}
and
\begin{equation}
\begin{array}{c|ccc|ccc|c|c|}
{} & {\alpha_0} & { \alpha_1} &
{ \alpha_2} & { f_0} & { f_1} &
{ f_2}& \sigma & z \\
\hline
{ \rho_0}&{ -\alpha_0} &{
 -\alpha_2} &{ -\alpha_1} &{
 f_0} &{ f_2} &{ f_1}&-\sigma &-z \\
\hline
{ \rho_1}&{ -\alpha_2} &{
 -\alpha_1} &{ -\alpha_0} &{
 f_2} &{ f_1} &{ f_0}&-\sigma &-z \\
\hline
{ \rho_2}&{ -\alpha_1} &{
 -\alpha_0} &{ -\alpha_2} &{
 f_1} &{ f_0} &{ f_2}&-\sigma &-z  
\end{array}
\label{rhoiss}
\end{equation}
that keep equations  \eqref{symp4} invariant.
The automorphisms $\pi_i$ and $\rho_i$ square to one
\begin{equation}
\pi_i^2 =1, \quad \rho_i^2=1, \quad i=0,1,2 \, ,
\label{pisquare}
\end{equation}
and satisfy 
the so-called braid relations 
\begin{equation}
\pi_i \pi_j \pi_i =\pi_j \pi_i \pi_j, \;\;\;
\rho_i \rho_j \rho_i =\rho_j \rho_i \rho_j, 
\quad i \ne j \, {.}
\label{braidrelations}
\end{equation}
The automorphisms $\pi_i$ and $\rho_i$ are related to automorphism 
$\pi$ from \eqref{p4symm} via
\begin{equation}
\pi= \pi_2 \pi_0= \pi_1\pi_2= \pi_0 \pi_1=\rho_2 \rho_0= \rho_1\rho_2= \rho_0 \rho_1
\label{braid1}
\end{equation}
and satisfy the following commutation
relations with the B\"acklund transformations $s_j$:
\begin{equation}
\pi_i s_i = s_i \pi_i, \quad \pi_i s_j = s_k \pi_i, 
\;\;\;
\rho_i s_i = s_i \rho_i, \quad \rho_i s_j = s_k \rho_i, 
\quad i \ne j, k
\ne j, i\ne k\,.
\label{braid4}
\end{equation}

We will now describe the B\"acklund transformations for the second order P$_{\rm IV}$
equations \eqref{p4}. %
The procedure will be illustrated 
by considering the $s_2$ transformation only. Generalizations to other
generators follow easily.

First, we consider $s_2 (\alpha_i), s_2 (f_i)$ from \eqref{piss} and
eliminate
 $\alpha_2=\sigma -\alpha_0-\alpha_1$ and $f_2=\sigma z-f_0-f_1$ to
obtain:
	\begin{align} \label{s2alpha2}
	s_2(\alpha_0)&=\sigma-\alpha_1,\qquad s_2(\alpha_1)=\sigma-\alpha_0,
      \\ 
      s_2(f_0)&=f_0+\frac{\sigma-\alpha_0-\alpha_1}{\sigma
      z-f_0-f_1}\label{s2f0}\\
      s_2(f_1)&=f_1-\frac{\sigma-\alpha_0-\alpha_1}{\sigma z-f_0-f_1}
      \, {.}
      \label{s2f1}
	\end{align}
Equation \eqref{noumisum} allows us to write down the following
relations between $f_0$ and $f_1$ :
	\begin{align}
	f_1&= \frac{-\alpha _0+ \sigma z f_0+f_0'-f_0^2}{2 f_0}\label{rel1}\\
	f_0&=\frac{\alpha _1+ \sigma z f_1-f_1'-f_1^2}{2 f_1}\label{rel2}
	\end{align}
used below to realize  $s_2$  as (1) B\"acklund and (2) auto-B\"acklund
transformations, respectively as shown below :
\begin{enumerate}
\item[(1)] Eliminating $f_0$ from the rhs of equation
	\eqref{s2f0} and $f_1$ from the rhs of equation
\eqref{s2f1}
	 yields:
	\begin{align}
		 s_2(f_0)&=\frac{2 f_1 \left(\sigma-\alpha _0-\alpha
	 _1\right)}{-\alpha _1+ \sigma z
	 f_1+f_1'-f_1{}^2}+\frac{\alpha _1+ \sigma z
	 f_1-f_1'-f_1{}^2}{2 f_1}, \label{s2f0f0}\\
	 s_2(f_1)&=-\frac{2 f_0 \left(\sigma-\alpha _0-\alpha
	 _1\right)}{\alpha _0+ \sigma z f_0-f_0'-f_0{}^2}-\frac{\alpha
	 _0- \sigma z f_0-f_0'+f_0{}^2}{2 f_0} \, .\label{s2f1f1}
\end{align}
\item[(2)] 
Inversely, eliminating $f_0$ from the rhs of equations
	\eqref{s2f1} and $f_1$ from the rhs of equations
and \eqref{s2f0}
	 yields  (note that in this case we denote $s_2$ by
	$\tilde{s}_2$ ):
		\begin{align}
	\tilde{s_2}(f_0)&=f_0+\frac{2 \left(\alpha _0 f_0+\alpha _1
	f_0-\sigma  f_0\right)}{-\alpha _0-\sigma  z
	f_0+f_0'+f_0{}^2},\label{ts2f0}\\
	\tilde{s_2}(f_1)&=f_1-\frac{2 \left(-\alpha _0 f_1-\alpha _1
	f_1+\sigma  f_1\right)}{-\alpha _1+\sigma  z
	f_1+f_1'-f_1{}^2}.\label{ts2f1}
	\end{align}
\end{enumerate}
Acting with $\rho_2$ connects relations \eqref{s2f0f0} and \eqref{s2f1f1}
as well as relations \eqref{ts2f0} and \eqref{ts2f1}:
\begin{equation}\label{rho2}
\rho_2 (s_2(f_0))
=s_2(f_1) ,\qquad \rho_2 (\tilde{s_2}(f_0)) =\tilde{s_2}(f_1)\, .
\end{equation}

As we saw above in items (1) and (2), $s_2(f_i), i=0,1$ could  either be
expressed in terms of $f_i , i=0,1$ or $f_j, i\neq j$ by simple
substitutions \eqref{rel1} or \eqref {rel2}. 
The transformation $s_2$ that maps  $f_0\to f_1$ and $f_1\to f_0$
is referred by us as \textbf{B\"acklund transformation} of
the system of second order P$_{\rm IV}$
equations \eqref{p4} and  maps equation \eqref{p4} with $i=0$ to that with
$i=1$ and vice versa.

The corresponding transformation that maps  $f_0\to f_0$ and $f_1\to f_1$
is denoted by as $\tilde{s_2}$ and is referred to as an \textbf{auto-B\"acklund
transformation} of the second order P$_{\rm IV}$
equation  \eqref{p4} with either $i=0$ or $i=1$. 

\section{Coalescence in the setting of symmetric P$_{\rm IV}$ equations} 
\label{section:coalescence}

In this section we look at coalescence 
in the setting of symmetric P$_{\rm IV}$ equations.
Such framework makes it easier to see what happens with the
B\"acklund symmetries in the $\epsilon \to 0$ limit.

Here we formulate coalescence in a setting of the  symmetric P$_{\rm
IV}$ equations \eqref{symp4}
through the following transformations :
\begin{equation}\label{te}
\begin{split}
f_i(z)&\to  f_i(z)+\frac{1}{\epsilon},\qquad z\to
z+\frac{2}{\sigma \epsilon^2},\\
\alpha_0&\to \epsilon \alpha_0 - \frac{1}{\epsilon^2},\qquad
\alpha_1\to \epsilon \alpha_1 + \frac{1}{\epsilon^2},\qquad
\alpha_2\to \epsilon \alpha_2 \, .
\end{split}
\end{equation}
Applying the above transformation to the first order equations \eqref{symp4} 
yields:
\begin{align}\label{p4coal}
f_0'(z)=&f_0 \left(f_1-f_2\right)+\frac{f_1-f_2}{\epsilon }+\epsilon \alpha_0 - \frac{1}{\epsilon^2}\nonumber\\
f_1'(z)=&f_1 \left(f_2-f_0\right)+\frac{f_2-f_0}{\epsilon }+\epsilon \alpha_1 + \frac{1}{\epsilon^2}\\
f_2'(z)=&f_2\left(f_0-f_1\right)+\frac{f_0-f_1}{\epsilon }+\epsilon \alpha_2\nonumber
\end{align}
Now we proceed by the same steps as in the preceding sections.
Summing the equations above we get:
\begin{equation*}
\epsilon\alpha_0+\epsilon\alpha_1+\epsilon\alpha_2=\epsilon
\sigma,\qquad f_0'+f_1'+f_2'=\epsilon \sigma.
\end{equation*}
Integrating equation 
$ \sum_i f_i'=\epsilon \sigma$ 
yields $ \sum_i f_i=\epsilon \sigma z +C$, where $C$ is an arbitrary constant of integration. 
Initially $C$ is set to zero but after applying transformation 
\eqref{te} on $f_i$ and $z$ we obtain :
\[
f_0+f_1+f_2+\frac{3}{\epsilon}=\epsilon \sigma z+\frac{2}{\epsilon}\, 
\;\; \longrightarrow f_0+f_1+f_2 = \epsilon \sigma
z-\frac{1}{\epsilon} \,.
\]
with $C=-1/\epsilon$.
Note that the presence of the non-zero integration constant does not affect the
symmetry of the symmetric  P$_{\rm IV}$ equations since we can always work 
with symmetry transformations acting on 
redefined $f_i$'s as will be done below.

Eliminating $f_2$ and $\alpha_2$ from \eqref{p4coal}, we get:
\begin{equation}\label{mixed}
\begin{split}
f_0'(z)&=\epsilon  \left(\alpha _0-\sigma z f_0\right)+\frac{2 f_0+2
f_1}{\epsilon }+f_0^2+2 f_1 f_0-\sigma z\,,\\
f_1'(z)&=\epsilon  \left(\alpha _1+\sigma z f_1\right)+\frac{-2 f_0-2
f_1}{\epsilon }-f_1^2-2 f_0 f_1+\sigma z \, .
\end{split}
\end{equation}
Substituting $\alpha_0=a_0/\epsilon, \alpha_1=a_1/\epsilon,
\sigma=\sigma_0/\epsilon$ with finite $a_0, a_1, \sigma_0$ and taking
$\epsilon \to \infty$ limit we recover  P$_{\rm IV}$ equations \eqref{noumisum}.

By eliminating $f_0$ from \eqref{mixed} we obtain:
\begin{equation}
\label{2ndeq}
\begin{split}
f_1''(z)=\frac{1}{\epsilon  f_1+1}&\left(\sigma -2 \alpha_0-2 \alpha_1-2 \sigma  z f_1+2 f_1{}^3+\epsilon ^3 \left(\frac{1}{2} \sigma ^2 z^2 f_1{}^2-\frac{\alpha_1^2}{2}\right)+\right.\\
&\epsilon ^2 \left(-2 \alpha_0 f_1{}^2-\alpha_1 f_1{}^2+\sigma ^2 z^2 f_1-2 \sigma  z f_1{}^3+\sigma  f_1{}^2\right)+\\
&\left.\epsilon  \left(-4 \alpha_0 f_1-2 \alpha_1 f_1-4 \sigma  z
f_1{}^2+2 \sigma  f_1+\frac{1}{2} f_1'{}^2+\frac{3}{2}
f_1{}^4+\frac{\sigma ^2 z^2}{2}\right)\right) {.}
\end{split}
\end{equation}
Taking instead the limit $\epsilon\to 0$ in equation \eqref{2ndeq} and
the corresponding equation for $f_0$ results in two copies of 
P$_{\rm II}$ equations, namely :
\begin{equation} \label{P2fi}
f_i''(z)=(-1)^i (-\sigma +2 \alpha_0+ 2\alpha_1)-2 \sigma  z f_i+2 f_i{}^3
\,,\quad i=0,1\, .
\end{equation}
The above P$_{\rm II}$ equations transform into each other under the
automorphism $\rho_2$ from \eqref{rhoiss}.
Since transformations \eqref{te} %
are nothing but M\"obius transformations on the variables $f_i$ and $z$, 
they  naturally preserve the Painlev\'e
property.

As a digression we note that equation \eqref{2ndeq} for $\sigma \to 0$ and 
finite $\epsilon$ becomes for $w=f_1-1/\epsilon$ : %
\begin{equation} \label{ince30}
w''(z)=\frac{w'^2}{2 w}+\frac{3 w^3}{2}-\frac{4 w^2}{\epsilon }-\frac{w \left(2 \alpha _0 \epsilon ^3+\alpha _1 \epsilon ^3-3\right)}{\epsilon ^2}-\frac{\left(\alpha _1 \epsilon ^3+1\right){}^2}{2 w \epsilon ^4}
\end{equation}
in which we recognize the equation XXX (I$_{30}$) of the Gambier's classification, that 
is listed in the classical book of  Ince \cite{ince} (see also \cite{inceshams}
for connection between Painlev\'e equations with additional parameters
and equations in \cite{ince}) as:
\begin{equation}\label{i30}
	I_{30}:\; w''(z)=\frac{w'^2}{2 w}+\frac{3 w^3}{2}+4 a w^2+2 b w+\frac{c}{w}\, .
\end{equation}
Also, if we make transformation $z\to z+\frac{2}{\sigma \epsilon^2} 
  -\xi/\sigma $ in equation \eqref{te} (equivalent to a different choice  of integration 
  constant $C$ in $\sum_i f_i=\sigma z+C$) with some new parameter $\xi$ and take the limit
$\epsilon \to 0$ in the corresponding second order equation for $f_0$ we
obtain
\begin{equation}
f_0''(z)=2 f_0^3-2( \sigma  z-\xi) f_0-\sigma +2 \alpha_0+ 2\alpha_1\,.
\end{equation}
 By taking $\sigma=0$ we arrive at 
Ince's I$_8$ equation:
	\begin{equation}\label{i8}
	I_{8}:\,\,w''=\,2 w^3+a w+b \,.
	\end{equation}

\subsection{The B\"acklund Transformations in the coalescence limit }

In this subsection we will show how $\mathcal{A}_2^{(1)}$ symmetry group   
reduces to  $\mathcal{A}_1^{(1)}$ symmetry in the
appropriate limit.

\subsubsection{$\mathcal{A}_2^{(1)}$ symmetry is maintained in
equations \eqref{p4coal} }\label{seca2sym}
Equations \eqref{p4coal} are invariant under:
\begin{equation} \label{a12coalescence}
\begin{array}{c|ccc|ccc|ccc}
{} & {\alpha_0} & { \alpha_1} &
{ \alpha_2} & { f_0} & { f_1} &
{ f_2}\\
\hline
{ s_0}&{ \frac{2}{\epsilon ^3}-\alpha _0} &{
	\alpha _0+\alpha _1-\frac{1}{\epsilon ^3}} &{ \alpha _0+\alpha _2-\frac{1}{\epsilon ^3}} &{
	f_0} &{ \frac{\alpha _0 \epsilon -\frac{1}{\epsilon ^2}}{f_0+\frac{1}{\epsilon }}+f_1} &{
	f_2-\frac{\alpha _0 \epsilon -\frac{1}{\epsilon ^2}}{f_0+\frac{1}{\epsilon }}}\\[2mm]
\hline
{ s_1}&{ \alpha _0+\alpha _1+\frac{1}{\epsilon ^3}} &{
	-\alpha _1-\frac{2}{\epsilon ^3}} &{ \alpha _1+\alpha _2+\frac{1}{\epsilon ^3}} &{
	f_0-\frac{\alpha _1 \epsilon +\frac{1}{\epsilon ^2}}{f_1+\frac{1}{\epsilon }}} &{ f_1} &{
	\frac{\alpha _1 \epsilon +\frac{1}{\epsilon ^2}}{f_1+\frac{1}{\epsilon }}+f_2} \\[2mm]
\hline
{ s_2}&{ \alpha_0+\alpha_2} &{
	\alpha_1+\alpha_2} &{-\alpha_2} &{
	\frac{\alpha _2 \epsilon }{f_2+\frac{1}{\epsilon }}+f_0} &{ f_1-\frac{\alpha _2 \epsilon }{f_2+\frac{1}{\epsilon }}} &{
	f_2} \\[2mm]
\hline
{ \pi}&{ \alpha_1+\frac{2}{\epsilon ^3}} &{
	\alpha_2-\frac{2}{\epsilon ^2}} &{ \alpha_0-\frac{2}{\epsilon ^2}} &{
	f_1} &{ f_2} &{ f_0} 
\end{array}
\end{equation}
and the automorphism $\rho_2$ from \eqref{rhoiss}.

After we eliminate $f_2$ and $\alpha_2$, we still have invariance under
$s_0, s_1$, but no longer under $\pi$ and the  $s_2$ transformation is modified to:
\begin{equation}
\begin{split}
s_2(f_0)=& \, f_0-\frac{\epsilon \left(-\alpha_0-\alpha _1+\sigma \right)}{f_0+f_1-\sigma  z \epsilon },\qquad s_2(f_1)=
f_1-\frac{\epsilon \left(-\alpha _0-\alpha _1+\sigma \right)}{-f_0-f_1+\sigma  z
\epsilon },\\
s_2(\alpha _0)=& \, \sigma -\alpha _1,
\qquad s_2(\alpha _1)= \sigma -\alpha _0 \, {.}
\end{split}
\label{s2f0f1}
\end{equation}

\subsubsection{Emergence of $\mathcal{A}_1^{(1)}$ symmetry in the $\epsilon \to0$ limit}
It is now easy to see from equation \eqref{a12coalescence} that the 
transformations $s_0,s_1$ and $\pi$ diverge in the limit $\epsilon\to 0$. 
Also $s_2$ becomes trivial in this limit. %
The way around this problem is to form 
the composition $s_0s_1s_0$ that will be shown not to
diverge in the limit $ \epsilon\to 0$
\cite{suzuki}. Similar 
ideas of  using compositions of B\"acklund transformations to 
obtain reduction from $A^{(1)}_l$ to $A^{(1)}_{l-k}$
appeared in  \cite{alred}.

The main conclusion of this subsection is 
that for the P$_{\rm IV}$ system of equations \eqref{p4coal}  for $f_0,f_1$ (obtained after
elimination of $f_2$) the $\epsilon \to
0$ limit will yield
 transformations $s_0s_1s_0$  (or identically  $s_1s_0s_1$) and 
$s_2$ as the two B\"acklund transformations that maintain P$_{\rm II}$ invariant.

Explicitly, the action of  $s_0s_1s_0$  on all variables is:
\begin{equation}\label{s0s2s0}
\begin{split}
s_0s_1s_0(f_0)&= f_0-\frac{\left(\alpha _0+\alpha _1\right) \epsilon 
\left(\epsilon f_0+1\right)}{\alpha _0 \epsilon ^2+\epsilon f_0
f_1+f_0+f_1}\, {,}\\
s_0s_1s_0(f_1)&= \frac{\left(\alpha _0+\alpha _1\right) \epsilon
^2}{\epsilon f_0+1}+\frac{\alpha _1^2 \epsilon ^4+\alpha _0 \alpha _1
\epsilon ^4+\alpha _0 \epsilon +\alpha _1 \epsilon }{\left(\epsilon 
f_0+1\right) \left(-\alpha _1 \epsilon ^2+\epsilon f_0
f_1+f_0+f_1\right)}+f_1\, {,}\\
s_0s_1s_0(\alpha_0)&=  -\alpha_1,\qquad s_0s_1s_0(\alpha_1)=-\alpha_0 \, {.}
\end{split}
\end{equation}
Now just looking at   transformations of the parameters
$\alpha_0,\alpha_1$ and using notation $\beta := \alpha_0+\alpha_1$ (since they are 
always together from now on), we see that they have a $A^{(1)}_1$ group 
structure due to:
\begin{equation}
\begin{array}{c|c|c}
{} & {\beta} & {\alpha_2}\\
\hline
{ s_0s_1s_0}&{-\beta} &{2\beta+\alpha_2}\\[2mm]
\hline
{ s_2}&{ 2\alpha_2+\beta} &{-\alpha_2} 
\end{array}
\end{equation}
and 
\[s_2^2=1,\qquad(s_0s_1s_0)^2=1,\qquad(s_0s_1s_0)s_2=s_2(s_0s_1s_0).\]

From now on we will use for brevity the following notation:
\begin{equation} \label{notation}
S_0 = s_0s_1s_0 , \qquad S_1= s_2 \, .
\end{equation} 
The relations \eqref{rel1}, \eqref{rel2} obtained  in section \eqref{BTsecond}
generalize to the following relations
\begin{equation}
\label{mix1}
\begin{split}
f_0=&\frac{\alpha _1 \epsilon ^2+\sigma  z \epsilon ^2 f_1-\epsilon  f_1'-
\epsilon  f_1{}^2-2 f_1+\sigma  z \epsilon }{2 \left(\epsilon  f_1+1\right)},\\
f_1=&\frac{-\alpha _0 \epsilon ^2+\sigma  z \epsilon ^2 f_0+\epsilon
f_0'-\epsilon  f_0{}^2-2 f_0+\sigma  z \epsilon }{2 \left(\epsilon
f_0+1\right)}\, ,
\end{split}
\end{equation} 
obtained from \eqref{mixed}.
Using relations \eqref{mix1} in an exactly the same way as we did
below equations 
\eqref{rel1}, \eqref{rel2}
we obtain two expressions for auto-B\"acklund transformations $\tilde{S_1}$ 
and B\"acklund transformations $S_1$ from those given in equation \eqref{s2f0f1}
\begin{gather*}
\tilde{S_1}(f_0)=f_0+ \frac{2 \left(-\alpha _0-\alpha _1+\sigma \right) \left(\epsilon  f_0+1\right)}{\alpha _0 \epsilon +\sigma z \epsilon  f_0-f_0'-f_0{}^2+\sigma z},\\
\tilde{S_1}(f_1)= f_1-\frac{2 \left(-\alpha _0-\alpha _1+\sigma \right) \left(\epsilon  f_1+1\right)}{-\alpha _1 \epsilon +\sigma z \epsilon  f_1+f_1'-f_1{}^2+\sigma z},\\
S_1(f_0)= \frac{\alpha _1 \epsilon ^2+f_1 \left(\sigma z \epsilon ^2-2\right)+\epsilon  \left(\sigma z-f_1'\right)-\epsilon  f_1{}^2}{2 \epsilon  f_1+2}+\frac{2 \left(-\alpha _0-\alpha _1+\sigma \right) \left(\epsilon  f_1+1\right)}{-\alpha _1 \epsilon +\sigma z \epsilon  f_1+f_1'-f_1{}^2+\sigma z},\\
S_1(f_1)= \frac{-\alpha _0 \epsilon ^2+f_0 \left(\sigma z \epsilon ^2-2\right)+\epsilon  \left(f_0'+\sigma z\right)+\epsilon  \left(-f_0{}^2\right)}{2 \epsilon  f_0+2}-\frac{2 \left(-\alpha _0-\alpha _1+\sigma \right) \left(\epsilon  f_0+1\right)}{\alpha _0 \epsilon +\sigma z \epsilon  f_0-f_0'-f_0{}^2+\sigma z}.
\end{gather*}
As in relations \eqref{rho2}, these two B\"acklund transformations $S_1$ and $\tilde{S_1}$  are related by 
the automorphism $\rho_2$.
Repeating the same steps for $S_0$ we
obtain :
\begin{gather*}
S_0(f_0)=  f_0-\frac{2 \left(\alpha _0+\alpha _1+\alpha _0 \epsilon  f_0+\alpha _1 \epsilon  f_0\right)}{\alpha _0 \epsilon +\sigma z \epsilon  f_0+f_0'-f_0{}^2+\sigma z},\\
\tilde{S_0}(f_0)= -\frac{2 \left(\alpha _0^2 \epsilon ^3+\alpha _0 \alpha _1 \epsilon ^3-\alpha _0-\alpha _1\right)}{\left(\epsilon  f_1+1\right) \left(-2 \alpha _0 \epsilon -\alpha _1 \epsilon -\sigma z \epsilon  f_1+f_1'+f_1{}^2-\sigma z\right)}+\\\frac{-2 \alpha _0 \epsilon ^2-\alpha _1 \epsilon ^2+\sigma z \epsilon ^2 f_1-\epsilon  f_1{}^2-2 f_1+\sigma z \epsilon }{2 \left(\epsilon  f_1+1\right)}-\frac{\epsilon  f_1'}{2 \left(\epsilon  f_1+1\right)}.
\end{gather*}
The B\"acklund transformations obtained in this way have non trivial limits for
$\epsilon\to 0$:
\begin{align}
\tilde{S_1}(f_0) =&\,  \frac{2 \left(-\alpha _0-\alpha
_1+\sigma \right)}{-f_0'-f_0{}^2+\sigma z}+f_0,&\tilde{S_1}(\beta)=
&2\sigma -\beta \, {,}
\label{S1P2}\\
S_1(f_0)=&\, \frac{2 \left(-\alpha _0-\alpha
_1+\sigma \right)}{f_1'-f_1{}^2+\sigma z}-f_1,&S_1(\beta)=
&2\sigma-\beta \, {,}\\
\tilde{S_0}(f_0)=& -\frac{2 \left(\alpha _0+\alpha
_1\right)}{f_0'-f_0{}^2+\sigma z}+f_0,&\tilde{S_0}(\beta)=&-\beta \, {,}
\label{S0tildef0}\\
S_0(f_0)=& -\frac{2 \left(-\alpha _0-\alpha
_1\right)}{f_1'+f_1{}^2-\sigma z}-f_1,&S_0(\beta)= &-\beta \, {.}
\label{S0P2}
\end{align}
These expressions agree with  B\"acklund transformations for P$_{\rm II}$ equation and they obey the 
$A^{(1)}_1$ group structure described in the literature \cite{gromak}\cite{kaji}
although the whole $A^{(1)}_1$ group structure requires presence of an
additional automorphism to be introduced below.

\subsubsection{The $\Pi$ automorphism for P$_{\rm II}$ model }

In this subsection we will construct automorphisms $\Pi, \widetilde{\Pi} $
of P$_{\rm II}$ equation that satisfy  $A^{(1)}_1$-type relations :
\begin{gather} \label{PPi}
\Pi(f_0)=f_1,\qquad \Pi(f_1)=f_0,\qquad \Pi(\beta)=\sigma-\beta,\\
\Pi S_i=S_j\Pi,\; i, j =0,1, \quad \qquad \Pi^2=1 \nonumber
\end{gather}
and
\begin{gather} \label{tPPi}
\widetilde{\Pi}(f_0)=-f_0,\qquad \widetilde{\Pi}(f_1)=-f_1,\qquad \widetilde{\Pi}(\beta)=\sigma-\beta\\
\widetilde{\Pi} \tilde{S_i}=\tilde{S_j}\widetilde{\Pi},\; i, j =0,1, \quad \qquad \widetilde{\Pi}^2=1
\, {,} \nonumber
\end{gather}
with $A^{(1)}_1$ transformations $S_i,\tilde{S_i}, i=0,1$ defined in equations
\eqref{S1P2}-\eqref{S0P2} as coalescence limits of appropriate $A^{(1)}_2$ 
transformations to be defined below.
Note that $f_i, i=0,1$ in the above relations satisfy P$_{\rm II}$ equations 
\eqref{P2fi}.

We now return to P$_{\rm IV}$ model where
we define $\mathcal{P}$ and $\mathcal{P}^{-1}$ : 
\begin{equation}
\mathcal{P}:=\pi s_0=s_1\pi,\qquad \mathcal{P}^{-1}:=s_0 \pi^2=\pi^2
s_1 \, {,}
\end{equation}
with $\pi $ and $s_i$ defined by relations \eqref{p4symm} 
from P$_{\rm IV}$ model.
The actions of $\mathcal{P}$ and $\mathcal{P}^{-1}$ on B\"acklund transformations $S_i$  
\eqref{notation} satisfy the following relations :
\begin{equation}
\begin{split} \label{mathcalPs}
S_0\mathcal{P}&=\mathcal{P}S_1:\qquad \{\alpha_0\to\alpha_0,\quad\alpha_1\to\alpha_1+\sigma\}
\, {,}\\
\mathcal{P}S_0&=S_1\mathcal{P}:\qquad \{\alpha_0\to\alpha_0-\sigma,\quad\alpha_1\to\alpha_1\}
\, {,}\\
S_0\mathcal{P}^{-1}&=\mathcal{P}^{-1}S_1:\qquad \{\alpha_0\to\alpha_0+\sigma,\quad\alpha_1\to\alpha_1\}
\, {,}\\
\mathcal{P}^{-1}S_0&=S_1\mathcal{P}^{-1}:\qquad
\{\alpha_0\to\alpha_0,\quad\alpha_1\to\alpha_1-\sigma\}\, .
\end{split}
\end{equation}
Accordingly $\mathcal{P}$ and $\mathcal{P}^{-1}$ satisfy the
product rules with $S_i$ identical to those given 
in relations \eqref{PPi} and \eqref{tPPi} 
although valid in the context of P$_{\rm IV}$ model.

Further one finds using the table \eqref{a12coalescence} and relations 
\eqref{mix1}
to calculate the actions $\mathcal{P}$ and $\mathcal{P}^{-1}$ on $f_i, i=0,1$
that they both converge to 
$\Pi$ and $\widetilde{\Pi}$ in the $\epsilon \to 0$ coalescence
limit. 
To illustrate this we will act with $\mathcal{P}$ on $f_i, i=0,1$ to
obtain according to the  table \eqref{a12coalescence} :
\begin{align}
\mathcal{P} ( f_0) &= \pi(f_0)=f_1 \, {,}\label{Pillustrate1}\\
\mathcal{P} ( f_1) &= \pi\left(f_1
+\frac{\alpha_0\epsilon-1/\epsilon^2}{f_0+1/\epsilon} \right) 
=f_2 +\frac{\alpha_1\epsilon+1/\epsilon^2}{f_1+1/\epsilon}\, {,}
\label{Pillustrate2}
\end{align}
where as we recall $f_2=\sigma \epsilon z -f_1-f_0-1/\epsilon$.
The relations \eqref{mix1} can now be used to substitute 
$f_1$ by 
$f_0$ on the right hand side of equation \eqref{Pillustrate1}
and $f_0$ by $f_1$ on the right hand side of equation
\eqref{Pillustrate2} giving in the limit $\epsilon \to 0$ the
result \eqref{tPPi}.  Using relation \eqref{mix1} to eliminate $f_1$
and substitute it by 
$f_0$ on the right hand side of equation
\eqref{Pillustrate2} gives in the limit $\epsilon \to 0$ the
result \eqref{PPi}.

\section{The mixed P$_{\rm II-IV}$ equations and its Hamiltonian}
\label{section:hamdeform}
We will  now consider the following class of generalizations of  P$_{\rm IV}$
equations  {\eqref{noumisum}} by adding nontrival terms parametrized by  constants 
$k_1,k_2$ :
\begin{align}
f_0^{\prime}  &= \alpha_0-\sigma zf_0 +f_0^2 +2 f_0f_1+\frac{1}{\epsilon}
\left(-k_1 \sigma z+k_2 (f_0+f_1)\right) \, {,}
\label{f0prbk}\\
f_1^{\prime}  &= \alpha_1+\sigma zf_1-f_1^2 -2 f_0f_1+\frac{1}{\epsilon}
\left(k_1 \sigma z-k_2 (f_0+f_1)\right) \, .
\label{f1prbk}
\end{align}
 We will determine values of 
constants  $k_1,k_2$  for which the above equations reproduce  P$_{\rm II}$ equation in the
$\epsilon \to 0$ limit. 

Note that we can write the equations
\eqref{f0prbk}-\eqref{f1prbk} as Hamilton equations 
with the Hamilton function \eqref{hamilton}, which generalized the
cubic P$_{\rm IV}$ Hamiltonian \eqref{okamotoP4} due to addition of 
quadratic terms with constants  $k_1,k_2$.

{First let us comment on how general are such extensions
of P$_{\rm IV}$ model.}
Replace the term $k_2 (f_0+f_1)$  on 
the right hand sides of equations \eqref{f0prbk} and \eqref{f1prbk}
with a more general  combination $k_2f_0+k_3f_1$
such that $k_2 \ne k_3$. In such case the resulting second order equation for 
$f_0$ and  $f_1$ would be divergent in the limit $\epsilon\to0$.
For example, $f_0^{\prime\prime}$ would contain the term
$(k_3-k_2)f_0^2/(2f_0 \epsilon^2+k_3 \epsilon)$ that would go to
infinity for $\epsilon \to 0$ unless $k_2=k_3$.
Thus, we have to set $k_2=k_3$ as we did in equations \eqref{f0prbk} and \eqref{f1prbk}. 
The addition of terms proportional to
$z f_i$ is also forbidden for the same reason. 

For $\alpha_i=a_i \epsilon,\,\, i=0,1,2 \,, \, \sigma=\sigma_0 \epsilon$ 
the second order equation for $f_0$ in the $\epsilon \to 0$ limit is:
\begin{equation}
f_0^{\prime\prime}= 2 f_0^3 +2 (k_1-k_2) \sigma_0 z f_0 + k_2 a_1+k_2 a_0
- k_1 \sigma_0 \, .
\label{P2bk}
\end{equation}
Thus as long as 
\begin{equation}
k_1 \ne k_2\, {,}
\label{bnek}
\end{equation}
the system \eqref{f0prbk}-\eqref{f1prbk} will have P$_{\rm II}$
equation as a limit.

We now discuss the conditions for the system
\eqref{f0prbk}-\eqref{f1prbk} to remain invariant under the $A^{(1)}_2$
symmetry.

Insert %
$f_2=\sigma z-f_1-f_0$
back into equations \eqref{f0prbk}, \eqref{f1prbk} and rewrite them as :
\begin{equation} \begin{split}
f_{0}'&=f_0\left( f_{1}-f_{2} \right) +\alpha_0 +\frac{1}{\epsilon}
(k_2(f_0+f_1)-k_1\sigma z)\, {,} \\
f_{1}'&=f_1\left( f_{2}-f_{0} \right) +\alpha_1-\frac{1}{\epsilon}
(k_2(f_0+f_1)-k_1\sigma z), \\
f_{2}'&=f_2\left( f_{0}-f_{1} \right) +\alpha_2 \, {,}
\end{split}\label{fffeps}
\end{equation}
with $\alpha_2=\sigma-\alpha_0-\alpha_1$.

Following Appendix \ref{section:integrconsts} we  now introduce
\begin{equation} %
{\bar f}_0=f_0+\frac{d}{\epsilon}, \qquad
{\bar f}_1=f_1+\frac{d}{\epsilon}, \qquad
{\bar f}_2=f_2
\label{fbarsd}
\end{equation}
in an effort to remove through this shift of $f_i$'s the extra terms 
with $k_1, k_2$ constants from the generalized P$_{\rm IV}$ equations
\eqref{f0prbk}-\eqref{f1prbk}.
In this way we obtain
\begin{equation} \begin{split}
{\bar f}_{0}'&={\bar f}_0\left( {\bar f}_{1}-{\bar f}_{2} \right)
+\alpha_0 
+\frac{d}{\epsilon^2} - \frac{d}{\epsilon}  
\left(  {\bar f}_{0} +{\bar f}_{1}-{\bar f}_{2} \right) 
+\frac{1}{\epsilon}
(k_2({\bar f}_0+{\bar f}_1-\frac{2d}{\epsilon})-k_1\sigma z) \, {,}
\\
{\bar f}_{1}'&={\bar f}_1\left( {\bar f}_{2}-{\bar f}_{0} \right)
+\alpha_1
-\frac{d}{\epsilon^2}
+ \frac{d}{\epsilon}  
\left(  {\bar f}_{0} +{\bar f}_{1}-{\bar f}_{2} \right) 
-\frac{1}{\epsilon}
(k_2({\bar f}_0+{\bar f}_1-\frac{2d}{\epsilon})-k_1\sigma z) 
, \\
{\bar f}_{2}'&={\bar f}_2\left( {\bar f}_{0}-{\bar f}_{1} \right)
+\alpha_2 \, {.}
\label{fbarepsd}
\end{split}
\end{equation}
In the first equation in \eqref{fbarepsd}  the terms with $({\bar
	f}_0+{\bar f}_1)$ and
the terms with $\sigma z$ will appear as
\begin{equation}
- \frac{(2 d -k_2)}{\epsilon} ({\bar f}_0+{\bar f}_1) + \frac{\sigma
	z}{\epsilon} (d-k_1)  \, {,}
\label{extraterms}
\end{equation}
after eliminating $f_2$ from this equation. The same terms but with the 
opposite sign will appear in the second equation in \eqref{fbarepsd}.

{With condition \eqref{bnek} satisfied we now describe two possible
cases, the first case coincides with the  P$_{\rm IV}$ 
coalescence model discussed in section \ref{section:coalescence} and
the second defines deformation of P$_{\rm IV}$ 
model to be discussed in section \ref{section:deformation}.}
\begin{enumerate}
\item[Case 1.] {Both terms in equation \eqref{extraterms} vanish.}
{This can only occur for 
\[ 2 d= k_2, \quad d=k_1\, ,
\]
which requires 
\begin{equation}
k_2= 2k_1\, .
\label{condiP4}
\end{equation}
Condition \eqref{condiP4} allows to restore 
the full $A^{(1)}_2$ symmetry in
the generalized P$_{\rm IV}$ equations
\eqref{f0prbk}-\eqref{f1prbk}.}
{Recall that such mechanism took place in the  P$_{\rm IV}$ 
coalescence model.
For example}, for $k_1=1,\,k_2=2,\,\sigma=\epsilon\sigma_0$ we recognize 
the coalescence
case of \eqref{mixed}.

\item[Case 2.] {Only one term in equation \eqref{extraterms} vanishes. 
Accordingly, we consider $k_2 \ne 2k_1$ and $k_1 \ne k_2$ (preserving
\eqref{bnek}).}
{Setting the variable $d$ to eliminate 
one of the two terms in \eqref{extraterms}, say
\[
d=k_1 ,
\]
results in $2 d -k_2= 2k_1 - k_2 \ne 0$. Consequently the only non-zero
extra term in the 
first equation in \eqref{fbarepsd} 
is  
\begin{equation} \label{2k1k2}
- \frac{(2 k_1 -k_2)}{\epsilon} ({\bar f}_0+{\bar f}_1) \, .
\end{equation}
Such system will be referred to as a deformed P$_{\rm IV}$  model
and will be discussed in the subsequent section.
One easily verifies that
choosing $d=k_2/2$ will result in a similar
model.}
\end{enumerate}

\section{Deformation of P$_{\rm IV}$ model}
\label{section:deformation}
As we have seen in section \ref{section:hamdeform}, P$_{\rm
II}$  equation can also be obtained from deformation of P$_{\rm IV}$ 
that changes its symmetry structure even before the limit is taken.

Following derivation presented in section \ref{section:hamdeform}
we now propose the following P$_{\rm IV}$  model :
\begin{equation}
{\bar H}=-f_0f_1f_2 
+
 \frac{-\alpha_1+\alpha_2}{3}f_0+\frac{-\alpha_1-2\alpha_2}{3}f_1 +
 \frac{2\alpha_1+\alpha_2}{3}f_2
+ \sum_{i,j,k}
\frac12 \eta_i (f_j+f_k)^2 \, ,
\label{H4defrom}
\end{equation}
as a generalization of the structure
in {\eqref{okamotoP4}}.
The summation in \eqref{H4defrom} is over all three indices 
$i,j,k$ being distinct.
The parameters $\eta_i, i=0,1,2$ are referred to as deformation parameters.

The corresponding equations are 
\begin{equation} \begin{split}
f_{0,z}&=f_0\left( f_{1}-f_{2} \right) +\alpha_0 -\eta_1(f_0+f_2) +
\eta_2 (f_0+f_1), \\
f_{1,z}&=f_1\left( f_{2}-f_{0} \right) +\alpha_1+\eta_0 (f_1+f_2)
-\eta_2 (f_0+f_1) , \\
f_{2,z}&=f_2\left( f_{0}-f_{1} \right) +\alpha_2 
-\eta_0 (f_1+f_2)+\eta_1(f_0+f_2)\, .
\end{split}\label{pfoureps012}
\end{equation}
Equations \eqref{pfoureps012} are invariant under automorphisms
\eqref{piss}, \eqref{rhoiss} augmented by
\[ \pi_i (\eta_i)=-\eta_i, \;\; \pi_i (\eta_j)=-\eta_k
\qquad  i, j, k \quad \text{distinct}
\]
and
\[ \rho_i (\eta_i)=\eta_i, \;\; \rho_i (\eta_j)=\eta_k
\qquad  i, j, k \quad \text{distinct} \, {.}
\]
Introduce
\begin{equation}
{\bar f}_i = f_i +\xi_i , \qquad i= 0,1,2 \, {,}
\label{fbari}
\end{equation}
with
\begin{equation}
\xi_i=\frac12 (\eta_j+\eta_k), \qquad  i, j, k \quad \text{distinct} \, {.}
\end{equation}
Note, that
\begin{equation}
\sum_i {\bar f}_i =  \sum f_i +\sum \xi_i
= \sigma z +\eta_0 +\eta_1 +\eta_2 \, {.}
\label{sumbfi}
\end{equation}
The equations \eqref{pfoureps012}  can then be recast back into the 
original form of P$_{\rm IV}$ symmetric equations:
\begin{equation} \begin{split}
{\bar f}_{0,z}&={\bar f}_0\left( {\bar f}_{1}-{\bar f}_{2} \right) +
{\bar \alpha}_0 \, {,}  \\
{\bar f}_{1,z}&={\bar f}_1\left( {\bar f}_{2}-{\bar f}_{0} \right) +
{\bar \alpha}_1 , \\
{\bar f}_{2,z}&={\bar f}_2\left( {\bar f}_{0}-{\bar f}_{1} \right) +{\bar \alpha}_2 
\, .
\label{pbfoureps012}
\end{split}
\end{equation}
but with the $z$-dependent coefficients:
\begin{equation} \begin{split}
{\bar \alpha}_0 &= \alpha_0 +\frac14 (\eta_1^2-\eta_2^2)
+\frac12 (\eta_2-\eta_1) \sigma z \, {,} \\
{\bar \alpha}_1 &= \alpha_1 +\frac14 (\eta_2^2-\eta_0^2)
+\frac12 (\eta_0-\eta_2) \sigma z  \, {,} \\
{\bar \alpha}_2 &= \alpha_2 +\frac14 (\eta_0^2-\eta_1^2)
+\frac12 (\eta_1-\eta_0) \sigma z  \, {,} 
\label{alphabar}
\end{split}
\end{equation}
that still satisfy $\sum {\bar \alpha}_i=\sum {\alpha}_i=\sigma$.

For $\eta_i=\eta_j, i\ne j$ the $z$-dependence will disappear from
${\bar \alpha}_k=\alpha_k, k \ne i, k \ne j$ and the system will become 
invariant under {\it one specific} B\"acklund transformation ${\bar s}_k$ defined as  one of the
following transformations:
\begin{equation}
\begin{array}{c|ccc|ccc|ccc}
{} & {{\bar \alpha}_0} & { {\bar \alpha}_1} &
{ {\bar \alpha}_2} & { {\bar f}_0} & { {\bar f}_1} &
{ {\bar f}_2}\\
\hline
{ {\bar s}_0}&{ -{\bar \alpha}_0} &{
	{\bar \alpha}_1+{\bar \alpha}_0} &{ {\bar \alpha}_2+{\bar \alpha}_0} &{
	{\bar f}_0} &{ {\bar f}_1+\frac{{\bar \alpha}_0}{{\bar f}_0}} &{
	{\bar f}_2-\frac{{\bar \alpha}_0}{{\bar f}_0}}\\[2mm]
\hline
{ {\bar s}_1}&{ {\bar \alpha}_0+{\bar \alpha}_1} &{
	-{\bar \alpha}_1} &{ {\bar \alpha}_2+{\bar \alpha}_1} &{
	{\bar f}_0-\frac{{\bar \alpha}_1}{{\bar f}_1}} &{ {\bar f}_1} &{
	{\bar f}_2+\frac{{\bar \alpha}_1}{{\bar f}_1}} \\[2mm]
\hline
{ {\bar s}_2}&{ {\bar \alpha}_0+{\bar \alpha}_2} &{
	{\bar \alpha}_1+{\bar \alpha}_2} &{-{\bar \alpha}_2} &{
	{\bar f}_0+\frac{{\bar \alpha}_2}{{\bar f}_2}} &{ {\bar f}_1-\frac{{\bar \alpha}_2}{{\bar f}_2}} &{
	{\bar f}_2} \\[2mm]
\end{array}  \, {.} 
\label{P4barsymmetries}
\end{equation}

Now set $\eta_0=\eta_1=0$ and $\eta_2= 2/\epsilon$ 
in \eqref{pfoureps012}. 
We see that in such case \eqref{pfoureps012} becomes \eqref{fffeps} with $k_1=0$ and $k_2=2$
and since $k_1 \ne k_2$ we know from equation \eqref{P2bk} that the
limit will still be P$_{\rm II}$.

The condition $\eta_i=\eta_j$ for $i\ne j$ and corresponding
invariance under $s_k$ transformation turns out to be
a condition for the model to pass Kovalevskaya-Painlev\'e test as 
we will now explain. 
\subsection{Kovalevskaya-Painlev\'e test of the Deformed Model \eqref{H4defrom}}

Assume that solutions of the extended P$_{\rm IV}$ \eqref{pfoureps012} equations
have the form 
\begin{equation}
f_i = \frac{a_i}{z} + b_i + c_i z + d_i z^2 + e_i z^3 + \cdots, \qquad
i=0,1,2 \, {.}
\label{fipoles}
\end{equation}
Substituting into \eqref{pfoureps012}  yields
\begin{align}
0&= a_i (a_{i+1}-a_{i-1})+a_i \label{KPeqs1} \, {,}\\
0&= a_i (b_{i+1}-b_{i-1})+b_i(a_{i+1}-a_{i-1})\nonumber 
\\&-\eta_{i+1}
(a_i+a_{i-1}) + \eta_{i-1} (a_i+a_{i+1})\label{KPeqs2} \, {,}\\
c_i&=a_i (c_{i+1}-c_{i-1})+\alpha_i +c_i (a_{i+1}-a_{i-1})
+b_i (b_{i+1}-b_{i-1}) \nonumber\\
&+\eta_{i-1} (b_i+b_{i+1})-\eta_{i+1}  (b_i+b_{i-1})\label{KPeqs3}\, {,}\\
2 d_i &= b_i (c_{i+1}-c_{i-1})+c_i (b_{i+1}-b_{i-1})+
a_i (d_{i+1}-d_{i-1}) \nonumber \\
&+ d_i  (a_{i+1}-a_{i-1}) +\eta_{i-1}(c_{i}+c_{i+1})
- \eta_{i+1}  (c_i+c_{i-1})  \, {,} \label{KPeqs4}
\end{align}
and etc for $i=0,1,2$.
Since $\sum_i a_i=0$ there are three (up to a sign and an overall
constant)  possible nontrivial solutions of the top
equation in \eqref{KPeqs1} 
\begin{align}
(a_0,a_1,a_2)&=(0,1,-1)\label{a0zero}\, {,}\\
(a_0,a_1,a_2)&=(-1,0,1)\label{a1zero}\, {,}\\
(a_0,a_1,a_2)&=(1,-1,0)\, ,\label{a2zero}
\end{align}
which correspond to $a_i=0$ for $i=0$ or $i=1$ or $i=2$. 
The automorphism $\pi_j$ will take the configuration with $a_i=0$
into the one with $a_k=0$ for the three distinct indices $i,j,k$.

We will show that 
for a given $i$ such that  
$a_i=0$ the solution \eqref{fipoles}
will pass the Kovalevskaya-Painlev\'e test 
\cite{Veselov-Shabat}
as long as
$\eta_j=\eta_k$.

We will illustrate the argument for $a_0=0$ as in \eqref{a0zero}. Plugging 
the sequence from \eqref{a0zero} into \eqref{KPeqs2} we find that 
\begin{equation}
b_0= - \frac12 (\eta_1+\eta_2), \quad b_2 =b_1 + \frac12
(\eta_2-\eta_1)  \, {.}
\label{bKPeqs}
\end{equation}
Thus, in the case of \eqref{a0zero} all the parameters $b_i$ are determined with exception of one,
either $b_1$ or $b_2$. For \eqref{a1zero} the
determined coefficient in term of $\eta$-coefficients 
will be $b_1$ with one
of $b_0$ or $b_2$ coefficients being undetermined.  For \eqref{a2zero} the
determined coefficient will be $b_2$ while one of the two other
coefficients remaining undetermined.
This is a general feature which is present independently of whether 
the $\eta$ deformation terms are present or not.

{}From \eqref{KPeqs3} we find that all the coefficients $c_i$ multiplying $z$ 
are determined
in terms of the lower coefficients:
\begin{align}
c_0&=-\alpha_0+(\eta_1+\eta_2)^2/4-\eta_1^2+b_1(\eta_1-\eta_2)\, {,}\\
c_1&=\frac13 (3\alpha_0+2\alpha_2+\alpha_1)+\frac13
b_1 (\eta_2-2\eta_0-\eta_1-b_1) +\frac16 \eta_1
(\eta_0-2\eta_2) \nonumber \\
&+(\eta_1^2-\eta_2^2)/4- \frac16 \eta_0 \eta_2
\, {,}\\
c_2&=\frac13 (3\alpha_0+2\alpha_1+\alpha_2)+\frac13
b_1 (2\eta_2+2\eta_0-2\eta_1+b_1)\nonumber\\&+\frac16
(\eta_0\eta_2 -\eta_1 \eta_2-
\eta_0\eta_1)
+ \eta_1^2/4  \, {.}
\end{align}
By summing the above coefficients one confirms that they satisfy the
condition
\begin{equation}
c_0+c_1+c_2=\alpha_0+\alpha_1+\alpha_2=\sigma  \, {,}
\label{cicondition}
\end{equation}
as expected from their definition in \eqref{fipoles}.

Let us rewrite equation \eqref{KPeqs4} as
\begin{equation}
\begin{split}
&2 d_i - d_i  (a_{i+1}-a_{i-1})- a_i (d_{i+1}-d_{i-1})
= b_i (c_{i+1}-c_{i-1})+c_i (b_{i+1}-b_{i-1})\\
&+\eta_{i-1}(c_{i}+c_{i+1})
- \eta_{i+1}  (c_i+c_{i-1})  \, {,}
\end{split}
\label{KPeqs4a}
\end{equation}
where we have grouped the terms with $d_i$ on the left hand side of the equation.
In all three \eqref{a0zero}, \eqref{a1zero} and \eqref{a0zero}  cases 
summing the left hand side of \eqref{KPeqs4a} 
over $i=0,1,2$ gives $2 (d_0+d_1+d_2)$
while the sum of the right hand side of \eqref{KPeqs4a} 
over $i=0,1,2$ vanishes as all the terms cancel
each other. This confirms that $\sum_i
d_i=0$ as expected from the definition in \eqref{fipoles}.

For the choice \eqref{a0zero} the left hand side of equation \eqref{KPeqs4a} vanishes
for $i=0$ while the right hand side is equal to 
\[ \frac12 (c_0+c_1+c_2) ( \eta_2-\eta_1)=
\frac12 \sigma ( \eta_2-\eta_1)  \, {.}
\]
Thus consistency requires in the case of \eqref{a0zero} that 
$ \eta_2=\eta_1$.
Similarly for the case \eqref{a1zero} we find  the left hand side of equation 
\eqref{KPeqs4a} vanishes
for $i=1$ while the right hand side is equal to $\sigma(
\eta_2-\eta_0)/2$ and for \eqref{a2zero} we find  the left hand side of equation 
\eqref{KPeqs4a} vanishes
for $i=2$ while the right hand side is equal to $\sigma (
\eta_1-\eta_0)/2$.  

Thus the condition for consistency is such that
$\eta_j=\eta_k$ for the case of $a_i=0$ with $d_i$ being the only
undetermined coefficient among $d_1,d_2,d_2$.
Generalizing the equation \eqref{KPeqs4a} to coefficient $f^{(k)}_i$ of
$z^k$ gives an equation with a left hand side:
$k f^{(k)}_i - f^{(k)}_i  (a_{i+1}-a_{i-1})- a_i (f^{(k)}_{i+1}-f^{(k)}_{i-1})
$. This relation can be cast in terms of the 
$3 \times 3$ matrix with a determinant $k(k-2)(2k+1)$. Correspondingly,
the undetermined coefficients only appear for $k=0$ and $k=2$ as one
of $b_i$ and $d_i$ coefficients consistent with what we have seen above. 
Together with a position of the pole
this leaves exactly three parameters as arbitrary with all the remaining
coefficients fully determined.
This demonstrates existence of a
solutions with simple pole structure and dependence on $3$ arbitrary constants
that are consistent when two of the deformations parameters are equal
to each other. 

Thus we have connected the integrability property associated with 
the fact of 
passing the Kovalevskaya-Painlev\'e test to presence of the B\"acklund 
symmetry under $s_i$  emerging from the consistency condition
$\eta_j=\eta_k$.

\subsection{ P$_{\rm II}$ limit of the deformed symmetric  P$_{\rm IV}$ equation}

The starting point here are equations %
\begin{equation} \begin{split}
 f_{0,z}&=f_0\left( f_{1}-f_{2} \right) +\alpha_0 +\eta (f_0+f_1), \\
 f_{1,z}&=f_1\left( f_{2}-f_{0} \right) +\alpha_1-\eta (f_0+f_1) , \\
 f_{2,z}&=f_2\left( f_{0}-f_{1} \right) +\alpha_2\, .
 \label{pfoureps2}
 \end{split}
 \end{equation}
 of the deformed P$_{\rm IV}$  obtained  from \eqref{pfoureps012} by 
setting $\eta_2=\eta, \eta_1=\eta_0=0$. The parameter $\eta$ is equal
to the constant $- (2 k_1 -k_2)/\epsilon$ in equation \eqref{2k1k2}
and as we have learned in section \ref{section:hamdeform}
equations \eqref{pfoureps2} will have P$_{\rm II}$ limit
which we elaborate in this section in greater details including 
application of the Painlev\'e test.

We recall that for $\eta_2=\eta, \eta_1=\eta_0=0$ equation
\eqref{pfoureps2}
is invariant under $s_2$ B\"acklund symmetry, $\pi_2$
automorphism from the  table \eqref{piss} with $\pi_2(\eta)=-\eta$ and
$\rho_2$ from the  table \eqref{rhoiss}  with $\rho_2(\eta)=\eta$ .

Using association $f_1=-q$ with $f_0+f_1+f_2= \sigma z$
we get from \eqref{pfoureps2} the following equation for $q$:
\begin{equation} \begin{split}
q_{zz}&= \frac{q_{z}^2}{2 q-\eta} +\frac{1}{2 q -\eta}
\left( 3 q^4 
 + 2q^3  ( 2\sigma z - \eta)+ 
q^2 (2\alpha_1+4\alpha_2-2 \sigma -5 \eta \sigma z
+ \sigma^2z^2 ) \right. \\
+&\left. q ( 3 \sigma \eta -2  \alpha_1 \eta -\eta \sigma^2 z^2 
+2 \eta^2 \sigma z -4 \alpha_2 \eta)-\sigma \eta^2
-\alpha_1^2 + \alpha_2\eta^2+ \eta \alpha_1 \sigma z  \right) \,
{.}
\label{qeP4}
 \end{split}
 \end{equation}

For $\eta\to 0$ we obtain P$_{\rm IV}$ equation :
\begin{equation} 
q_{zz}= \frac{q_{z}^2}{2 q} +\frac{3 q^3}{2 q }
 + 2q^2 \sigma z+ q (\alpha_1+2\alpha_2- \sigma + \frac12 \sigma^2z^2 ) 
 -\frac{\alpha_1^2}{2q} {,}
\label{rP4}
 \end{equation}
{that agrees with equation \eqref{p4} for $f_1=-q$.}  %

For $\sigma \to 0$ and $Q=q-\eta/2$:
\begin{equation} \begin{split}
Q_{zz}&= \frac{Q_{z}^2}{2 Q} +\frac32 Q^3
 +2 Q^2   \eta+ 
Q \left(\alpha_1+2\alpha_2 +\frac{3}{4} \eta^2  \right) \\
&-\frac{1}{2Q}
\left(\alpha_1 +\frac{1}{4}
\eta^2 \right)^2 \,{,}
\label{QP4r0}
 \end{split}
 \end{equation}
which is 
I$_{\rm 30}$ for $\eta \ne 0$.

For $\sigma=\sigma_0/\eta, \alpha_1=a_1/\eta, \alpha_2=a_2/\eta$
and in the limit $\eta \to \infty$ we get :
\begin{equation}
q_{zz}= 2 q^3 -2 q \sigma_0 z- a_2+\sigma_0= 2 q^3 -2 q \sigma_0 z + (a_0+a_1)
 \, {.}
\label{PIIeq}
\end{equation}
More generally for $f_0$ and $f_1$ from equation \eqref{pfoureps2} 
we obtain in the limit $\eta \to \infty$ ;
\begin{equation} \label{P2afi}
f_i''(z)=(-1)^i ( \alpha_0+ \alpha_1)-2 \sigma_0  z f_i+2 f_i{}^3
\,,\quad i=0,1\, .
\end{equation}
in which we recognize  two P$_{\rm II}$ equations for $i=0$ and $i=1$
that again are transformed 
into each other under the automorphism $\rho_2$ from \eqref{rhoiss}
but differ from P$_{\rm II}$ equations in  \eqref{P2fi} by the values of
the constant coefficients on the right hand sides.

Because of the presence 
of deformation parameter $ \eta$ in the denominator in relation  \eqref{qeP4} 
it appears 
that the three cases $\eta \ll 1, \eta \gg 1$ and
$\eta$-finite need to be considered separately. 
For the first two cases we are in  P$_{\rm IV}$ and P$_{\rm II}$
regimes, respectively
but for finite $\eta$ it makes sense to make a change of variables
$q \to Q=q-\eta/2$ with corresponding equation
\begin{equation} \begin{split}
Q_{zz}&= \frac{Q_{z}^2}{2 Q} +\frac32 Q^3
 +2 Q^2  (\sigma z+ \eta)+ 
Q \left(\frac12 \sigma^2 z^2 + \frac12 \eta \sigma z 
+ \alpha_1+2\alpha_2 -\sigma +\frac{3}{4} \eta^2  \right) \\
&+\frac12 \sigma \eta  +\frac{1}{2Q}
\left(  \eta \sigma z \alpha_1 - \alpha_1^2 - \frac{1}{2}
\eta^2 \alpha_1 - \frac14 \eta^2 \sigma^2 z^2 + \frac14 \eta^3
\sigma z - \frac{1}{16} \eta^4\right) {.}
\label{QP4}
 \end{split}
 \end{equation}
In Appendix \ref{painlevetest} we 
provide details of the Painlev\'e test applied on equation
\eqref{QP4}.  That equation
\eqref{QP4} passes the direct Painlev\'e test agrees with 
the result of the {Kovalevskaya- Painlev\'e}  test that established the
consistency of the extended P$_{\rm IV}$ \eqref{pfoureps012}  as long as 
two out three $\eta_i$ parameters are equal (which is the case here).

\subsection{First order P$_{\rm II}$ equations as a limit of the deformed model}
Here we will show how starting from equations \eqref{pfoureps2} to
obtain the first order system of equations
underlying the P$_{\rm II}$ equations \eqref{P2afi}  and their 
$A^{(1)}_1$ B\"acklund transformations in a limit $\eta \to \infty$.
We set %
 $\sigma=\sigma_0/\eta, \alpha_1=a_1/\eta, \alpha_2=a_2/\eta$
with constants $\sigma_0, a_1, a_2$ 
and represent $f_1, f_2$ as 
\begin{equation}
f_1= - q , \quad f_2=- \frac{2}{\eta} p, \quad 
f_0 = \frac{\sigma_0}{\eta} z + \frac{2}{\eta} p +q \, .
\label{fp2subs}
\end{equation}
Plugging these substitutions into \eqref{pfoureps2} %
we obtain
\begin{align}
- \frac{2}{\eta} p_z&= - \frac{2}{\eta} p \left(
 \frac{\sigma_0}{\eta} z +\frac{2}{\eta} p +2 q
\right)+\frac{a_2}{\eta} \, {,}\label{p2epsilonpz}\\
-q_z &= -q  \left( - \frac{\sigma_0}{\eta} z -q- \frac{4}{\eta} p \right)
+\frac{a_1}{\eta}-\eta \left( \frac{\sigma_0}{\eta} z +
\frac{2}{\eta} p 
\right) \, . \label{p2epsilonqz}
\end{align}
Considering large $ \eta$ and neglecting the terms of 
order $O(1/\eta^2)$ in the first equation and 
the terms of order $O(1/\eta)$  one obtains in such limit
equations
\begin{align}
 p_z&= 2 pq - \frac12 a_2 \, {,} \label{p2pz}\\
q_z &= -q^2 +\sigma_0z +2 p\, {.}  \label{p2qz}
\end{align}
Taking the derivative with respect to $z$ on both sides of
\eqref{p2qz} gives P$_{\rm II}$
equation  \eqref{PIIeq}  (or \eqref{P2afi} with $f_1=-q$).

Let us now repeat the above analysis to obtain  the P$_{\rm II}$
equation \eqref{P2afi}  with a different sign of the constant term. 
We consider
\begin{equation}
f_0= - y , \quad f_2=- \frac{2}{\eta} h, \quad 
f_1 = \frac{\sigma_0}{\eta} z + \frac{2}{\eta} h +y \, .
\label{fp2asubs}
\end{equation}
that follows from identificiation \eqref{fp2subs} via acting
with $\rho_2$ automorphism and replacing $q,p$ with $y,h$
to emphasize that we are working with a different P$_{\rm II}$
equation. Plugging  substitutions \eqref{fp2asubs}
into \eqref{pfoureps2} like in \eqref{p2epsilonqz}
and considering large $ \eta$ we arrive at the system of first order
equations :
\begin{align}
h_z&= -2 h y - \frac12 a_2  \, {,} \label{p2apz}\\
y_z &= y^2 -\sigma_0z -2 h \, {,} \label{p2aqz}
\end{align}
that lead to the second P$_{\rm II}$
equation namely \eqref{P2afi} with $f_0=-y$.

{}From equation \eqref{p2qz} we derive
\begin{equation} 
p= \frac12 (q_z+q^2-\sigma_0 z ) \, {.} 
\label{pdef}
\end{equation} 
In order to conveniently introduce all the $A^{(1)}_1$ symmetry generators
in the setting of equations \eqref{p2pz}, \eqref{p2qz}
let us define an auxiliary quantity $v$ obtained from $p$ given in 
\eqref{p2qz} by transformation $q \to -q$ :
\begin{equation} 
v= \frac12 (-q_z+q^2-\sigma_0 z )= -p+q^2 - \sigma_0 z \, {.}
\label{vdef}
\end{equation} 
Taking a derivative on both sides of equation \eqref{vdef} we obtain the
counterparts of equations \eqref{p2pz}, \eqref{p2qz} valid for 
$v,q$:
\begin{align}
 v_z&= -2 vq + \frac12 (-\sigma_0 - a_0-a_1)  \, {,} \label{v2vz}\\
q_z &= q^2 -\sigma_0z -2 v \, {.} \label{v2qz}
\end{align}
Note that the transformation :
\begin{equation} 
\overline{\Pi} : q\to -q , \quad p \to v , \quad  \sigma_0 \to
\sigma_0 , \quad
a_0+a_1\to - a_0-a_1 \, {,}
\label{cappi}
\end{equation} 
takes equations \eqref{p2pz}, \eqref{p2qz} into equations 
\eqref{v2vz}, \eqref{v2qz} and does not change P$_{\rm II}$
equation  \eqref{PIIeq}.
Also note that the transformation :
\begin{equation} 
{\bar \rho} : z \to -z , \quad \sigma_0 \to -\sigma_0 , \quad p \to v , \quad 
q\to q \, ,
\label{caprho}
\end{equation} 
will have the same effect. 

 Alternatively, equations \eqref{v2vz}, \eqref{v2qz} can be obtained
 directly
 from symmetric deformed P$_{\rm IV}$ equations \eqref{pfoureps2}
 through the following substitution of $f_1, f_2$ :
 \begin{equation} \label{f1f2v}
  f_1=  q , \quad f_2= \frac{2}{\eta} v, \quad 
 f_0 = \frac{\sigma_0}{\eta} z - \frac{2}{\eta} v -q, \quad a_2=
 -(\sigma_0+a_0+a_1) \, {,}
 \end{equation} 
 for large $ \eta$ values.

Recall that for the deformed P$_{\rm IV}$  equations \eqref{pfoureps012}  
with $\eta=\eta_2\ne 0$ and $\eta_1=\eta_0=0$ the 
surviving symmetry generator is $s_2$:
\begin{equation} \label{s2f1f2}
f_1 \stackrel{s_2}{\longrightarrow}  f_1 - \frac{\alpha_2}{f_2} , \qquad f_2  \stackrel{s_2}{\longrightarrow} 
f_2 ,\;\;
\quad \alpha_2   \stackrel{s_2}{\longrightarrow}  - \alpha_2 \, {,}
\end{equation} 
or in terms of variables $p,q$ used above :
\begin{equation} 
q  \stackrel{s_2}{\longrightarrow}  q - \frac{a_2}{2p} , \qquad p  \stackrel{s_2}{\longrightarrow}  p ,\;\;\quad 
a_2   \stackrel{s_2}{\longrightarrow}  - a_2 \, {,}
\label{s2pa}
\end{equation} 
after cancellation of $\eta$. One easily checks that indeed
equations \eqref{p2pz}, \eqref{p2qz} are invariant under $s_2$
transformation as shown in \eqref{s2pa}. 
Note that $s_2: a_0+a_1 \to -(a_0+a_1)+2\sigma_0$.

Similarly inserting representation \eqref{f1f2v} into expression for
the $s_2$-transformation \eqref{s2f1f2} produces after cancellation of $\eta$
\begin{equation} 
q  \stackrel{\tilde{s}_2}{\longrightarrow}  q + \frac{\sigma_0+a_0+a_1}{2v} , \quad 
v \stackrel{\tilde{s}_2}{\longrightarrow}  v ,\;\;\;
a_0 + a_1  \stackrel{\tilde{s}_2}{\longrightarrow}  - (a_0+a_1)-2\sigma_0 \, ,
\label{s2va}
\end{equation} 
where we denoted $s_2$ by $\tilde{s}_2$ when it acts on $q,v$ system to
distinguish it from $s_2$ as defined in relations \eqref{s2pa}.
Both transformations $s_2$ and $\tilde{s}_2$ defined in 
\eqref{s2pa} and \eqref{s2va} keep the P$_{\rm II}$  equation
\eqref{PIIeq} invariant and square to one. 
It is interesting to compare action of $\tilde{s}_2$ to that of 
the automorphism $\tilde{S_0}=s_0s_1s_0$ from equation \eqref{S0tildef0}.
Introducing $\gamma= \sigma_0+a_0+a_1$ we can rewrite the nontrivial part
of transformation \eqref{s2va} as 
\begin{equation} 
q  \stackrel{\tilde{s}_2}{\longrightarrow}  q + \frac{\gamma}{-q_z+q^2-\sigma_0 z  } , 
\;\;\; \gamma 
 \stackrel{\tilde{s}_2}{\longrightarrow}  - \gamma\, ,
\label{s2vaa}
\end{equation} 
where we inserted the definition of $v$ from equation \eqref{vdef}.
Comparing with expression \eqref{S0tildef0} we see that the action of
$\tilde{s}_2$ almost agree with the limit of $s_0s_1s_0$ and the difference is
only due to the difference between constant terms of P$_{\rm II}$ equations
given in \eqref{P2afi} versus \eqref{P2fi}.

Using relation \eqref{vdef} between $v$ and $p$ one also derives
formulas for actions of $\tilde{s}_2$ on $p$ and $s_2$ on $v$ :
\begin{align} 
v &\stackrel{s_2}{\longrightarrow} v - \frac{q}{p} a_2 +
\frac{a_2^2}{4p^2} \, ,
\label{s2v}\\
p &\stackrel{\tilde{s}_2}{\longrightarrow}  p + \frac{q}{v}(\sigma_0+a_0+a_1)
+ \frac{(\sigma_0+a_0+a_1)^2}{4v^2}\, .
\label{st2p}
\end{align} 
These completes all the information on  the B\"acklund transformations
of the $A^{(1)}_1$ symmetry group consisted of $s_2,\tilde{s}_2,
\bar{\Pi}$ 
of P$_{\rm II}$ equation obtained as a limit
of the deformed  P$_{\rm IV}$ model. 

Note that equations \eqref{p2pz}, \eqref{p2qz} and equations 
\eqref{v2vz}, \eqref{v2qz} can be compactly summarized as
a system of equations
\begin{equation} \begin{split} \label{summary1order}
q_z&=p-v \, {,}\\
 p_z&= 2 pq - \frac12 a_2 \, {,} \\
v_z&= -2 vq + \frac12 (-\sigma_0 - a_0-a_1)  \, {.}
\end{split}
\end{equation} 
manifestly invariant under $\bar{\Pi}$ and $\bar{\rho}$.

Similarly,  equations \eqref{p2pz}, \eqref{p2qz} would enter into a 
system of equations
\begin{equation} \begin{split} \label{summary1aorder}
y_z&=u-h \, {,}\\
h_z&= -2 hy - \frac12 a_2  \, {,} \\
u_z&= 2 uy + \frac12 (-\sigma_0 - a_0-a_1)  \, {.}
\end{split}
\end{equation} 
that lead to the other copy of  P$_{\rm II}$
equations in  \eqref{P2afi} with its own $A^{(1)}_1$ symmetry.

\section{Concluding comments}
\label{section:concluding}
We would like to make few comments on special novel features of our 
formalism.

By enlarging a parameter space of P$_{\rm IV}$ model we extended the
$A^{(1)}_2$ symmetry structure by additional automorphisms
$\pi_i,\rho_i, i=0,1,2$.
In particular, the presence of the automorphism 
$\rho_2$  facilitated the reduction process from $A^{(1)}_2$ to $A^{(1)}_1$.
The authomorphism $\rho_2$ together with the B\"acklund transfomation $s_2$ 
remain a symmetry for 
P$_{\rm II-IV}$ and survive the P$_{\rm II}$ limit while they also commute with each
other. {
A  crucial feature of  P$_{\rm II}$ limit  of P$_{\rm IV}$ generalized
models 
is that it consists of two P$_{\rm II}$ equations (see \eqref{P2fi},
\eqref{p2pz},\eqref{p2qz} and \eqref{p2apz},\eqref{p2aqz})
connected via authomorphism $\rho_2$. Each of the two P$_{\rm II}$ 
equations is invariant under $A^{(1)}_1$ symmetry.
Thus the presence of $\rho_2$ is critically important for
the full understanding of  all features of the formalism. 
Note that in order to define the action of authomorphisms  
$\pi_i,\rho_i, i=0,1,2$ they need to be formulated
on an enlarged parameter space that includes 
$\sigma$ 
that transforms nontrivially under 
these authomorphisms, see tables \eqref{piss}, \eqref{rhoiss}.
The presence of $\sigma$ affords us also an opportunity to include 
in the formalism the solvable Painlev\'e equations (classified by Gambier) 
that appear on Ince's list  \cite{ince} (see also \cite{inceshams}). 
In particular, equations I$_{30}$, I$_{8}$ given in equations
\eqref{i30},\eqref{i8} were obtained here in the $\sigma \to 0$ limit.
}

{
As long as $\sigma$ remains non-zero there exists 
a transformation 
\[
\alpha_i=\tilde{\alpha_i} \sigma,\qquad f_i(z)=
\sqrt{\sigma}\tilde{f_i}(\tilde{z}),\qquad z=\tilde{z}/\sqrt{\sigma}
\]
given in terms of quantities entering equations \eqref{symp4} that
allows for absorbing $\sigma$ in the formalism and thus 
effectively setting it to $1$.
Note however that the possibility of redefining $\sigma=\sigma_0/\eta$
is essential for ability of taking  P$_{\rm II}$  limit for  $\eta \to \infty $ 
in the deformed P$_{\rm IV}$ model. Likewise, setting 
$\sigma=\sigma_0/\epsilon$ and taking
$\epsilon \to \infty$ limit was crucial for recovering  P$_{\rm IV}$ equations  
from equation \eqref{mixed}. Since we are interested in models that
interpolate between  P$_{\rm II}$  and P$_{\rm IV}$ and in
automorphisms $\pi_i,\rho_i, i=0,1,2$ we  
work here with the formalism depending explicitly on $\sigma$.
}

As pointed out below the definition of the  Hamiltonian $H$ in \eqref{hamilton} 
the choice of additional terms in $H$ ensured invariance under $\rho_2$. As explained 
below \eqref{hamilton}, equivalent
theories with invariance under $\rho_0$ or  $\rho_1$ could be 
introduced via simple redefinitions of the additional terms.
We have provided arguments that there is one unique (up to simple
redefinition of such additional terms)
generalization of P$_{\rm IV}$ model %
allowing addition of quadratic terms to the hamiltonian and requiring
finite limits and Painlev\'e property.

To summarize, in this work we focused on the symmetry properties for the
two generalizations, namely coalescence and deformation, of P$_{\rm IV}$
model contained in P$_{\rm II-IV}$ Hamiltonian of \eqref{hamilton}.
We derived B\"acklund transformations of the P$_{\rm II-IV}$ model 
and uncovered a connection between presence of symmetry and 
passing of Painlev\'e property test. This work raises an interesting 
question whether other Painlev\'e/Ince equations can be 
unified within  some mixed model similar to 
the one presented in this paper.

\appendix

\section{ About introducing two integration constants into the P$_{\rm IV}$ system}
\label{section:integrconsts}
Above, we have studied the transformation \eqref{fbarsd} with $\sum_i {\bar
	f}_i= \sum_i f_i +2 d /\eta$. We would therefore now investigate
the P$_{\rm IV}$ systems that generally allow for $ \sum_i f_i= \sigma
z +C$.

Given is the P$_{\rm IV}$ system 
\begin{equation}
f_i^{\prime}= f_i \left( f_{i+1}-f_{i+2} \right) +\alpha_i, \quad i=0,1,2
\label{P4std}
\end{equation}
invariant under B\"acklund symmetries $s_i, i=0,1,2$ and $\pi$.
The two constraints of the P$_{\rm IV}$ system 
\begin{equation}
\sum_i \alpha_i=\sigma, \qquad \sum_i f_i=\sigma z+C
\label{P4intcst}
\end{equation}
define two possible integration constants $\sigma,C$ of the P$_{\rm IV}$ system.
Customarily, people set $C=0$ and $\sigma=1$. Recall that setting 
$\sigma=0$ reduces P$_{\rm IV}$ to Ince's 
XXX equation (see also \cite{inceshams}).

The integration constant $C$ can be absorbed by redefining $f_i$'s :
$f_i \to g_i$ so that $ \sum_i g_i=\sigma z $ and the system is obviously
still invariant  under B\"acklund symmetries $s_i, i=0,1,2$ and $\pi$.

There appear (at least) three ways of changing variables to eliminate
$C$ from the constraint $ \sum_i f_i$. In each of these cases, the
constant  $C$ will appear
explicitly in the resulting differential equations.
\begin{enumerate} 
	\item 
	\begin{equation}
	g_i = f_i +\eta ,\;\; i=0,1,2 , \quad  3 \eta =-C
	\label{redef1}
	\end{equation}
	with the shifted P$_{\rm IV}$ system :
	\begin{equation}
	g_i^{\prime} = g_i \left( g_{i+1}-g_{i+2} \right) +\alpha_i -\eta 
	\left( g_{i+1}-g_{i+2} \right) 
	, \quad i=0,1,2
	\label{P4redef1}
	\end{equation}
	\item 
	\begin{equation}
	h_i = f_i +\eta ,\;\; i=0,1 , \; h_2= f_2, \quad  2 \eta =-C
	\label{redef2}
	\end{equation}
	with the shifted P$_{\rm IV}$ system :
	\begin{equation}
	\begin{split}
	h_{0,z}&=h_0\left( h_{1}-f_{2} \right) +\alpha_0+\eta^2-
	\eta (h_0+h_1-f_2), \\
	h_{1,z}&=h_1\left( f_{2}-h_{0} \right) +\alpha_1-\eta^2 -\eta 
	(f_2 -h_0-h_1)\\
	f_{2,z}&=f_2 \left( h_0-h_1  \right) +\alpha_2
	\end{split}
	\end{equation}
	with $ f_2 =\sigma z -h_0-h_1- 2 \eta = \sigma z -h_0-h_1+C$.
	\item 
	\begin{equation}
	d_0 = f_0 +\eta ,  \quad  \eta =-C
	\end{equation}
	with the shifted P$_{\rm IV}$ system :
	\begin{equation}
	\begin{split}
	d_{0,z}&=d_0\left( f_{1}-f_{2} \right) +\alpha_0-
	\eta (f_1-f_2), \\
	f_{1,z}&=f_1\left( f_{2}-d_{0} \right) +\alpha_1+\eta f_1
	\\
	f_{2,z}&=f_2 \left( d_0-f_1  \right) +\alpha_2 - \eta f_2 \,.
	\end{split}
	\label{P4redef3}
	\end{equation}
\end{enumerate} 
We refer the reader to section \ref{section:hamdeform} where the above scheme 2.
was employed to transform accordingly generalized P$_{\rm IV}$ equations.

\section{ Painlev\'e test of equation \eqref{QP4} } 
\label{painlevetest}
In this appendix  we will apply the Painlev\'e test to  equation \eqref{QP4}. 
Following the standard procedure of this test we first insert 
\[ Q(z)= a_0 (z-z_0)^\mu
\]
and  focus on the dominant behavior near singularity
on both sides of equation
\eqref{QP4} to obtain 
\[ \mu(\mu-1) a_0 (z-z_0)^{\mu-2} = \frac{a_0^2 \mu^2  
(z-z_0)^{2\mu-2}}{ 2 a_0 (z-z_0)^\mu}+\frac{3}{2} a_0^3 (z-z_0)^{3 \mu} 
\]
with contributions on the right hand side originating from the first
and the  second
term of the right hand side of  equation
\eqref{QP4}. 
This way we obtain: 
\[ a_0^2 =1 ,\qquad\quad \mu=-1 
\]
consistent with the Painlev\'e requirement that $\mu$ is a negative integer
for a movable pole with no branching.
Next, to check the resonance condition  we plug
\[ Q (z) = a_0 (z-z_0)^{-1} + \eta (z-z_0)^{-1+R}
\]
into  equation \eqref{QP4} and keep only the terms linear in $\eta$ to
obtain the resonance equation for $R$:
\[  (R+1)(R-3)=0
\]
This resonance structure suggests that a Laurent expansion
\begin{equation}
q (z) = \sum_{j=0}^{\infty} a_j  (z-z_0)^{j-1}=
a_0 (z-z_0)^{-1} + a_1 + a_2 (z-z_0)+ h (z-z_0)^{2} + a_4 (z-z_0)^{3}
+ \cdots
\label{Laurent}
\end{equation}
expresses expansion around an arbitrary pole at $z_0$
where we identified $a_3=h$ as the single arbitrary coefficient.
Inserting expression \eqref{Laurent} into \eqref{QP4} and looking on
coefficients of power of $\eta=z-z_0$ we get:
\begin{equation} \begin{split}
 0&=-a_0^2+a_0^4 \\
0&=3 a_0^3 a_1 +a_0^3 \eta-a_0a_1 +a_0^3 \sigma z_0\\
0&=12 a_0^2 a_1 \sigma z_0+12 a_0^3 a_2-6 a_2 a_0+ a_0^2 \eta \sigma z_0+3
a_0^2 \eta^2/2+4 a_0^3 \sigma+4 a_0^2 \alpha_2\\
&-2 a_0^2 \sigma+18 a_0^2 a_1^2+
2 a_0^2 \alpha_1+ a_0^2 \sigma^2 z_0^2+12 a_0^2 a_1 \eta\\
0&= \sigma \eta a_0+12 a_0 a_1^2 \sigma z_0+8 a_0 a_1 \alpha_2+12 a_2 a_0^2 \sigma z_0+
12 a_1^3 a_0\\
&+12 a_0^2 a_1 \sigma+12 a_3 a_0^3+36 a_2 a_0^2 a_1+2 a_0 a_1 z_0 \eta \sigma+
3 a_0 a_1 \eta^2\\
&+4 a_0 a_1 \alpha_1+12 a_0 a_1^2 \eta-4 a_0 a_1 \sigma+
2 a_0^2 \sigma^2 z_0-12 a_3 a_0+a_0^2 \eta \sigma\\
&+2 a_0 a_1 \sigma^2 z_0^2
+12 a_2 a_0^2 \eta
\label{pwrexp}
\end{split}
 \end{equation}
The top equation gives two possible non-zero solutions
\[ a_0= \pm 1
\]
The second equation gives:
\begin{equation}
a_1= - \frac12 (\eta+\sigma z_0)
\label{a1b}
\end{equation}
for both values $a_0=1$ and $a_0=-1$.

The third equation gives two values for $a_2$:
\begin{equation}
a_2 = \frac{\eta \sigma  z_0}{3} +\frac{\sigma^2 z_0^2}{12} - \frac{2
\alpha_2}{3} -\frac{\sigma}{3}- \frac{\alpha_1}{3}
\label{a2cp1}
\end{equation}
for  $a_0=1$ and $a_1$ as given in \eqref{a1b}
and
\begin{equation}
a_2 = -\frac{\eta \sigma z_0}{3} -\frac{\sigma^2 z_0^2}{12} + \frac{2
\alpha_2}{3} -\sigma +\frac{\alpha_1}{3}
\label{a2cm1}
\end{equation}
for $a_0=-1$ and $a_1$ as given in \eqref{a1b}.

Consider now the fourth equation. The coefficient $a_3$ drops from
this equation for both values of $a_0=\pm 1$ as it should (resonance $R=3$).
The solutions of the the fourth equation for $a_2$ are
\begin{equation}
a_2 = \frac{\eta \sigma z_0}{3} +\frac{\sigma^2 z_0^2}{12} - \frac{2
\alpha_2}{3} -\frac{\sigma}{3}- \frac{\alpha_1}{3}
\label{a2dp1}
\end{equation}
for  $a_0=1$ and $a_1$ as given in \eqref{a1b}
and
\begin{equation}\begin{split}
a_2&= \frac{-1}{12(\eta+ \sigma z_0)} ( 
5 \sigma^2 z_0^2 \eta+\sigma^3 z_0^3+16 \sigma \eta+
12 \sigma^2 z_0\\
&+4 \sigma z_0 \eta^2-4 \alpha_1 \eta-4 \alpha_1 \sigma z_0-
8 \alpha_2 \eta-8 \alpha_2 \sigma z_0 )
\end{split}
\label{a2dm1}
\end{equation}
for $a_0=-1$ and $a_1$  as given in \eqref{a1b}.
We see that $a_2$ given in \eqref{a2cp1} and \eqref{a2dp1} are equal so the
solution to the recursive problem is in this case consistent. In addition
$a_4$ and higher coefficients will depend on $a_3$ but $a_3$ is not fixed 
by the scheme and can be taken to any value including zero.

However $a_2$ given in \eqref{a2cm1} and \eqref{a2dm1} differ  by
$\sigma
\eta/(3(\sigma z_0+\eta))$. Hence  the second solution for $a_0=-1$ 
is only consistent for $\eta=0, \sigma \ne 0$  or $\sigma=0, \eta
\ne 0$.

It seems therefore that as long as $a_0=1$ 
there is a solution to the recurrence
relations that does not fail the  Painlev\'e test.

\vspace{5mm}
{\bf Acknowledgments}
JFG and AHZ thank CNPq and FAPESP for financial support. VCCA 
thanks S\~ao Paulo Research Foundation (FAPESP) for financial support 
by grants 2016/22122-9 and 2019/03092-0.

\end{document}